\documentclass[superscriptaddress,showpacs,twocolumn,nofootinbib]{revtex4-2}

\usepackage{sgmath}
\usepackage{booktabs}
\usepackage{makecell}
\usepackage{graphicx}
\usepackage{color}
\usepackage{subfigure}
\usepackage{amssymb}
\usepackage{amsmath}
\usepackage{bm}
\usepackage{mathtools}
\usepackage{ragged2e}
\usepackage{hyperref}
\hypersetup{
    colorlinks=true,
    citecolor=MidnightBlue,
    linkcolor=BrickRed,
    urlcolor=black,
    breaklinks=true,
}
\usepackage[dvipsnames]{xcolor}

\newcommand{\tightmatrix}[1]{%
  \begingroup
  \renewcommand{\arraystretch}{0.85}%
  $\begin{bmatrix}#1\end{bmatrix}$%
  \endgroup
}

\begin{document}

\title{Anomalous tensorial properties of anisotropic 2D materials}

\author{Elizabeth J. Dresselhaus}
\email{Correspondence: ej$_$dresselhaus@berkeley.edu, s$_$g@berkeley.edu, kranthi@berkeley.edu\\}
\affiliation{University of California, Berkeley, California 94720, USA}

\author{Sanjay Govindjee}
\email{Correspondence: ej$_$dresselhaus@berkeley.edu, s$_$g@berkeley.edu, kranthi@berkeley.edu\\}
\affiliation{University of California, Berkeley, California 94720, USA}

\author{Kranthi K. Mandadapu}
\email{Correspondence: ej$_$dresselhaus@berkeley.edu, s$_$g@berkeley.edu, kranthi@berkeley.edu\\}
\affiliation{University of California, Berkeley, California 94720, USA}
\affiliation{Lawrence Berkeley National Laboratory, California 94720, USA}

\begin{abstract}
Odd transport phenomena---defined as a flux response orthogonal to an applied gradient---have been recently observed in isotropic systems, with a multitude of proposed models and experiments to study these effects. Odd transport manifests in tensors that describe linear relations between fluxes and gradients that drive them, particularly when parity and time-reversal symmetries are broken. 
In this work, we identify such odd properties to be a subset of a broader class of major-symmetry-breaking behaviors, which we term ``anomalous." 
We develop a classification of anomalous properties described by $2^\mathrm{nd}$ and $4^\mathrm{th}$ order tensors in anisotropic 2D materials that maintain discrete rotational and reflection symmetries, characterized by the 17 wallpaper groups.
To this end, we present representation theorems for these tensors, identifying which components are constrained for specific spatial symmetries and thereby allowing materials to be grouped into classes that exhibit anomalous responses or not.
We focus our discussion on $2^\mathrm{nd}$ order tensors in the context of electrical resistivity and on $4^\mathrm{th}$ order tensors in the context of viscosity and elasticity. 
These findings are broadly applicable to the study of novel emergent material properties. To illustrate this, we discuss implications of our findings for two very different 2D materials that have recently garnered attention in condensed matter physics: knitted fabrics and twisted bilayer graphene. 
\end{abstract}

\maketitle

\vspace{0.1in}
\noindent\textbf{\textit{Introduction.}} In conventional materials, elastic deformations store energy and viscous phenomena dissipate energy. In $d \geq 2$ dimensions, the linear response coefficients for elasticity and viscosity are tensorial quantities. Thus, odd elasticity, which does not store energy, and odd viscosity, which does not dissipate energy, are possible in 2D materials~\cite{fruchart-review}. 
In a similar spirit, electric current conventionally flows parallel to an applied electric field. However, in 2D electron gases, applying an out-of-plane magnetic field deflects electrons perpendicular to the electric field, leading to odd contributions to the conductivity---the so-called Hall effect~\cite{tongQHE}.

While the Hall effect is observed in most metals and semiconductors \cite{karsenty2020comprehensive}, only certain 2D materials exhibit odd linear response coefficients in elasticity and viscosity. 
Recent experiments on active matter systems composed of spinning starfish embryos \cite{tan2022odd} and specially engineered robotic metamaterials \cite{brandenbourger2019nonreciprocal} have shown evidence for odd elasticity \cite{scheibner2020odd}.  
Odd viscosity \cite{Avr98, Banerjee2017, kranthi-odd-viscosity, Soni2019, Hargus2020}, on the other hand, can arise in isotropic systems if and only if those systems are two-dimensional and break both time-reversal and parity symmetries \cite{kranthi-odd-viscosity}. Recent work has also uncovered the possibility of odd viscodiffusive fluids, which exhibit cross-coupling between viscous and diffusive behaviors \cite{deshpande2024odd}. Theoretical works thus far have focused on finding odd phenomena in isotropic systems \cite{scheibner2020odd, Avr98, Banerjee2017, kranthi-odd-viscosity, Soni2019,Hargus2020,Hargus2021,  deshpande2024odd,fruchart-review, fruchart2022odd}. While many physical systems can be well-approximated as isotropic, the rapidly growing field of metamaterials now allows for  the systematic and tunable design of anisotropic materials \cite{MechMetaMatRev}. Moreover, odd phenomena represent only a special case of a broader class of behaviors arising from major symmetry breaking.\footnote{For second-order material property tensors, major symmetry breaking implies $T_{ij}\not = T_{ji}$ and in the fourth-order case that $T_{ijkl}\not=T_{klij}$.} This leaves open the identification and classification of other anomalous responses resulting from such symmetry breaking. Defining any tensor that breaks major symmetry to be \emph{anomalous}, in this work, we predict the possibility of anomalous phenomena in anisotropic materials that retain varying degrees of spatial symmetry.
We do so by developing general representation theorems for tensors of different orders that describe material behaviors in the context of electrical resistivity, viscosity and elasticity.

\begin{table*}[htbp]
\caption{Wallpaper groups classified by point group symmetry, with tensor structures for second- and fourth-order responses. Extra spacing separates distinct point groups for visual clarity.  $\ddag$ superscripts indicate cases where symmetry considerations permit anomalous behavior.\\}
\label{t:wallpaper_tensor}
\centering
\renewcommand{\arraystretch}{1.6}
\setlength{\tabcolsep}{8pt}
\resizebox{\linewidth}{!}{
\begin{tabular}{
>{\raggedright\arraybackslash}p{2.6cm}
>{\raggedright\arraybackslash}p{2.6cm}
>{\raggedright\arraybackslash}p{5.2cm}
>{\centering\arraybackslash}m{4.8cm}
>{\centering\arraybackslash}m{4.0cm}
}
\toprule
\textbf{Hermann-Mauguin notation} & \textbf{Orbifold signature} & \textbf{Point groups and elements} & \textbf{$4^\mathrm{th}$ order tensor structure} & \textbf{$2^\mathrm{nd}$ order tensor structure} \\
\midrule

p1 & o & \makecell[l]{\hspace{1em}$\mathit{C}_1$:\quad$\mathbb{I}$} &
\tightmatrix{\eta_{1111} & \eta_{1122} & \eta_{1112} \\ \eta_{2211} & \eta_{2222} & \eta_{2212} \\ \eta_{1211} & \eta_{1222} & \eta_{1212}}\textsuperscript{\ddag} &
\tightmatrix{\rho_{11} & \rho_{12} \\ \rho_{21} & \rho_{22}}\textsuperscript{\ddag} \\
\addlinespace[8pt]

p2 & 2222 & \makecell[l]{\hspace{1em}$\mathit{C}_2$:\quad$\mathbb{I}$, $-\mathbb{I}$} &
\tightmatrix{\eta_{1111} & \eta_{1122} & \eta_{1112} \\ \eta_{2211} & \eta_{2222} & \eta_{2212} \\ \eta_{1211} & \eta_{1222} & \eta_{1212}}\textsuperscript{\ddag} &
\tightmatrix{\rho_{11} & \rho_{12} \\ \rho_{21} & \rho_{22}}\textsuperscript{\ddag} \\
\addlinespace[10pt]

\makecell[l]{pm, pg, cm} & \makecell[l]{**, XX, *X} & \makecell[l]{\hspace{1em}$\mathit{D}_1$:\quad$\mathbb{I}$, $B_0$} &
\tightmatrix{\eta_{1111} & \eta_{1122} & 0 \\ \eta_{2211} & \eta_{2222} & 0 \\ 0 & 0 & \eta_{1212}}\textsuperscript{\ddag} &
\tightmatrix{\rho_{11} & 0 \\ 0 & \rho_{22}} \\
\addlinespace[10pt]

\makecell[l]{p2mm, p2mg,\\ p2gg, c2mm} & \makecell[l]{*2222, 22*,\\ 22X, 2*22} & \makecell[l]{\hspace{1em}$\mathit{D}_2$:\quad$\mathbb{I}$, $-\mathbb{I}$, $B_0$, $B_\pi$} &
\tightmatrix{\eta_{1111} & \eta_{1122} & 0 \\ \eta_{2211} & \eta_{2222} & 0 \\ 0 & 0 & \eta_{1212}}\textsuperscript{\ddag} &
\tightmatrix{\rho_{11} & 0 \\ 0 & \rho_{22}} \\
\addlinespace[10pt]

p3 & 333 & \makecell[l]{\hspace{1em}$\mathit{C}_3$:\quad$\mathbb{I}$, $A_{2\pi/3}$, $A_{4\pi/3}$} &
\tightmatrix{\eta_{1111} & \eta_{1122} & \eta_{1112} \\ \eta_{1122} & \eta_{1111} & -\eta_{1112} \\ -\eta_{1112} & \eta_{1112} & \tfrac{1}{2}(\eta_{1111} - \eta_{1122})}\textsuperscript{\ddag} &
\tightmatrix{\rho_{11} & \rho_{12} \\ -\rho_{12} & \rho_{11}}\textsuperscript{\ddag} \\
\addlinespace[10pt]

\makecell[l]{p3m1, p31m} & \makecell[l]{*333, 3*3} & \makecell[l]{\hspace{1em}$\mathit{D}_3$:\quad$\mathbb{I}$, $A_{2\pi/3}$, $A_{4\pi/3}$,\\ \hspace{4em}$B_{\pi/3}$, $B_\pi$, $B_{5\pi/3}$\\
 \hspace{3.5em}($B_0$, $B_{2\pi/3}$, $B_{4\pi/3}$)} &
\tightmatrix{\eta_{1111} & \eta_{1122} & 0 \\ \eta_{1122} & \eta_{1111} & 0 \\ 0 & 0 & \tfrac{1}{2}(\eta_{1111} - \eta_{1122})} &
\tightmatrix{\rho_{11} & 0 \\ 0 & \rho_{11}} \\
\addlinespace[6pt]


p4 & 442 & \makecell[l]{\hspace{1em}$\mathit{C}_4$:\quad$\mathbb{I}$, $A_{\pi/2}$,\\ \hspace{4em}$-\mathbb{I}$, $A_{3\pi/2}$} &
\tightmatrix{\eta_{1111} & \eta_{1122} & \eta_{1112} \\ \eta_{1122} & \eta_{1111} & -\eta_{1112} \\ \eta_{1211} & -\eta_{1211} & \eta_{1212}}\textsuperscript{\ddag} 
&
\tightmatrix{\rho_{11} & \rho_{12} \\ -\rho_{12} & \rho_{11}}\textsuperscript{\ddag} \\
\addlinespace[10pt]

\makecell[l]{p4mm, p4gm} & \makecell[l]{*442, 4*2} & \makecell[l]{\hspace{1em}$\mathit{D}_4$:\quad$\mathbb{I}$, $A_{\pi/2}$, $-\mathbb{I}$, $A_{3\pi/2}$,\\ \hspace{4em}$B_0$, $B_{\pi/2}$, $B_\pi$, $B_{3\pi/2}$} &
\tightmatrix{\eta_{1111} & \eta_{1122} & 0 \\ \eta_{1122} & \eta_{1111} & 0 \\ 0 & 0 & \eta_{1212}} &
\tightmatrix{\rho_{11} & 0 \\ 0 & \rho_{11}} \\
\addlinespace[10pt]

p6 & 632 & \makecell[l]{\hspace{1em}$\mathit{C}_6$:\quad$\mathbb{I}$, $A_{\pi/3}$, $A_{2\pi/3}$,\\ \hspace{4em}$-\mathbb{I}$, $A_{4\pi/3}$, $A_{5\pi/3}$} &
\tightmatrix{\eta_{1111} & \eta_{1122} & \eta_{1112} \\ \eta_{1122} & \eta_{1111} & -\eta_{1112} \\ -\eta_{1112} & \eta_{1112} & \tfrac{1}{2}(\eta_{1111} - \eta_{1122})}\textsuperscript{\ddag} &
\tightmatrix{\rho_{11} & \rho_{12} \\ -\rho_{12} & \rho_{11}}\textsuperscript{\ddag} \\
\addlinespace[10pt]

p6mm & *632 & \makecell[l]{\hspace{1em}$\mathit{D}_6$:\quad$\mathbb{I}$, $A_{\pi/3}$, $A_{2\pi/3}$, $-\mathbb{I}$,\\ \hspace{4em}$A_{4\pi/3}$, $A_{5\pi/3}$, $B_0$, $B_{\pi/3}$,\\ \hspace{4em}$B_{2\pi/3}$, $B_\pi$, $B_{4\pi/3}$, $B_{5\pi/3}$} &
\tightmatrix{\eta_{1111} & \eta_{1122} & 0 \\ \eta_{1122} & \eta_{1111} & 0 \\ 0 & 0 & \tfrac{1}{2}(\eta_{1111} - \eta_{1122})} &
\tightmatrix{\rho_{11} & 0 \\ 0 & \rho_{11}} \\

\bottomrule
\end{tabular}}
\end{table*}


\vspace{0.1in}
\noindent\textbf{\textit{Wallpaper groups, orbifold signatures and point groups.}} To classify such anomalous responses, we consider the group of distance-preserving maps of the 2D plane. This group has two connected components: the set of all rotations parametrized by angle $\theta$, denoted by $A_{\theta}$, and the set of reflections about a line at polar angle $\alpha/2$ with respect to $x-$axis, denoted by $B_{\alpha}$---in addition to translations. As matrices, $A_\theta$ and $B_\alpha$ are of the form: 
\begin{equation}
    A_\theta = \begin{pmatrix} \cos\theta & -\sin\theta \\ \sin\theta & \cos\theta \end{pmatrix}\, , \ 
    B_\alpha = \begin{pmatrix} \cos\alpha & \sin\alpha \\ \sin\alpha & -\cos\alpha \end{pmatrix}\,.
\end{equation}
Combinations of these maps can tile the plane and be categorized into 17 distinct \emph{wallpaper groups} based on the orders of rotation centers in the materials and the presence or absence of reflection symmetries \cite{schwarzenberger:1974, schattschneider:1978,armstrong:1988}. 
Each wallpaper group may contain a combination of (a) reflections, (b) glide reflections, and (c) rotation symmetries. 
The different wallpaper groups are generally denoted via their orbifold signature~\cite{sym_of_things} or via IUCr (Hermann-Mauguin) notation~\cite{ITC_Vol_A}.

Every wallpaper group is associated with a point group, which is either a cyclic group or a dihedral group of order $n$ ($C_n$ or $D_n$) \cite{ITC_Vol_A}. Multiple wallpaper groups can correspond to a single point group; they differ in whether or not rotation centers lie on mirror axes and in the relative orientations of mirror and glide axes.\footnote{Throughout this work, we use the terms ``mirror" and ``reflection" interchangeably.} From the perspective of coarse-grained continuum modeling, it is the point group symmetries that control the structure of macroscopic material properties. The 17 wallpaper groups with their IUCr and orbifold designations are shown in Columns I and II of Table \ref{t:wallpaper_tensor}.  They are grouped into equivalence classes defined by their corresponding point groups\footnote{For equivalence class $D_3$ we provide two common but equivalent sets of reflection elements.} as shown in Column III.

\vspace{0.1in}
\noindent\textbf{\textit{Constitutive relations as tensorial maps.}} 
Coarse-grained material properties and responses to external fields are commonly described by constitutive laws that relate “fluxes” (${J}$) to the corresponding “forces” (${X}$), typically expressed as a functional relation: ${J} = {f}({X})$. In the context of mechanics, the force corresponds to a displacement or velocity gradient, with the response given by a stress tensor, whereas in the context of mass or charge transport, the force is a gradient of a density or electric potential and the response is a mass flux or electric current. These functions may, in general, be non-linear; however, for small gradients, they are often well-approximated by linear maps of the form:
${J} = {L}{X}$~\cite{Groot1984}, where ${L}$ is a material property tensor of the appropriate order, determined by the tensorial order of ${J}$ and ${X}$.
The interrelations among elements of the tensor ${L}$, including how many independent components are required to specify it, and what anomalies are present, can be deduced from symmetries inherent in the material.  

Consider, for instance, material properties described by second-order tensors, such as conductivity, which relates the electric field to the electric current density, and diffusivity, which maps a concentration gradient to a diffusive flux. Consider specifically the conductivity tensor and its inverse, the resistivity tensor ${\rho}$, we may write, ${E} = {\rho} {J}$, where ${E}$ and ${J}$ denote the electric field and current density vectors, respectively. The major symmetry condition $\rho_{ij} = \rho_{ji}$ may then be invoked, reducing the number of independent components of ${\rho}$. This symmetry reflects the Onsager reciprocal relations~\cite{Onsager1931a,Onsager1931b}, which arise from time-reversal symmetry of the underlying microscopic dynamics. For strictly isotropic systems, the resistivity tensor is typically proportional to the identity, i.e., a diagonal tensor with equal elements. However, when a magnetic field is applied perpendicular to the plane of a conducting material, electrons are deflected transversely by the Lorentz force, inducing a Hall voltage orthogonal to the applied current. This effect is captured by a non-zero off-diagonal term in the resistivity tensor. Under isotropy, the representation theorem for tensors in 2D requires the off-diagonal elements to satisfy $\rho_{21} = -\rho_{12}$~\cite{tongQHE, kranthi-odd-viscosity}, exemplifying odd transport—a subclass of anomalous transport characterized by major symmetry breaking. Fundamentally, this behavior originates from the breaking of time-reversal symmetry by the magnetic field. A similar odd response appears in systems of active rotary particles that break microscopic time-reversal symmetry and parity, where diffusive fluxes emerge perpendicular to concentration gradients~\cite{Hargus2021,hargus2025flux}.

Another instructive example involves $4^\text{th}$ order tensors arising from constitutive relations between observables described by symmetric second-order tensors.
Here, we focus on the viscosity tensor $\eta_{ijkl}$, which couples symmetric stresses $\sigma_{ij}$ to symmetric velocity gradients $d_{ij}$ (strain-rates) via Newton’s law: $\sigma_{ij} = \eta_{ijkl} d_{kl}$.
In 2D, the full $2^4 = 16$ components of $\eta_{ijkl}$ can be reduced even without invoking any material symmetries.
Since both stress and strain-rate tensors are symmetric, we impose the so-called minor symmetries: $\eta_{ijkl} = \eta_{jikl} = \eta_{ijl k}$,
leaving 9 independent components.
As in the case of second-order tensors, microscopic time-reversal symmetry imposes an additional constraint via the Onsager reciprocal relations. This condition, major symmetry, requires $\eta_{ijkl} = \eta_{kl ij}$~\cite{kranthi-odd-viscosity}. 
For isotropic systems, these symmetries reduce $\eta_{ijkl}$ to just two independent coefficients: the bulk and shear viscosities that characterize a typical Newtonian fluid.
However, when time-reversal symmetry is broken, major symmetry no longer holds, and an additional, antisymmetric component of the viscosity tensor is allowed:
$\eta_{1112} = - \eta_{1211} =-\eta_{2212} =  \eta_{1222}= \mu^{\mathrm{o}}$
~\cite{Avr98, Banerjee2017, kranthi-odd-viscosity, fruchart-review,Soni2019}.
This $\mu^{\mathrm{o}}$, known as odd viscosity, describes a response where shear induces a normal stress and is dissipation-free.
An analogous breaking of major symmetry has been reported in the context of linear elasticity under non-conservative internal forces, leading to odd elasticity~\cite{odd-elasticity}. 

While anomalous transport in isotropic systems is well-characterized by classical representation theorems for isotropic tensors~\cite{weyl1946classical,jeffreys1973isotropic,kranthi-odd-viscosity}, it remains unclear which anomalous behaviors are permitted under the 17 wallpaper group symmetries listed in Table~\ref{t:wallpaper_tensor}.
In what follows, we systematically develop the corresponding representation theorems for second- and fourth-order tensors for each equivalence class of wallpaper groups.

\vspace{0.1in}
\noindent\textbf{\textit{Symmetry classification of anomalous transport: Second-order tensors.}} 
Using resistivity as a canonical example, the representation of $2^\mathrm{nd}$ order tensors can be analyzed via the invariance requirement:
\begin{equation}\label{eq:2nd-transformation}
    \rho_{ij} = Q_{ip} Q_{jq} \rho_{pq}
\end{equation}
for all $Q_{ij}$ in the point group associated with a wallpaper group. 

The implications of major symmetry breaking can be understood by separately considering reflection and rotation symmetries. For instance, consider a reflection (or glide reflection) symmetry with the reflection axis, without loss of generality, aligned along the $x$-axis, corresponding to ${Q} = B_0$. In this case, Eq.~\eqref{eq:2nd-transformation} leaves the diagonal elements unconstrained, but requires the off-diagonal elements to vanish. Thus, no major symmetry-breaking anomalies arise for material behaviors described by second-order property tensors when $B_0$ is a symmetry element.
By contrast, consider a material with a rotational symmetry $Q = A_\theta$. Applying Eq.~\eqref{eq:2nd-transformation} yields equal diagonal components, $\rho_{11} = \rho_{22}$, and antisymmetric off-diagonal components, $\rho_{12} = -\rho_{21}$. This condition holds for all rotation angles $\theta$ excepting when  
$\theta = \pi$, a two-fold rotational symmetry that imposes no constraint on the tensor elements.

In summary, anomalous transport, associated with major symmetry breaking, can only occur in wallpaper groups that lack  reflection symmetries.
Applying Eq.~\eqref{eq:2nd-transformation} to the point groups associated with the 17 wallpaper groups yields the representation theorems summarized in Column~V of  Table~\ref{t:wallpaper_tensor}; see Appendix~\ref{app:B} for detailed derivations. 
Among these, only the groups p1~(o), p2~(2222), p3~(333), p4~(442), and p6~(632) exhibit major symmetry breaking anomalous properties. Of these, p1~(o) and p2~(2222) support anisotropic anomalies, while p3~(333), p4~(442), and p6~(632) exhibit anomalies analogous to those found in odd transport in isotropic systems.  

\vspace{0.1in}
\noindent\textbf{\textit{Symmetry classification of anomalous transport: Fourth-order tensors.}} Anomalies in material properties represented by $4^\mathrm{th}$ order tensors can be analyzed using the invariance requirement~\cite[\S 7.2]{anandgovindjee}
\begin{equation}\label{eq:4thorder-transform}
    \eta_{ijkl} = Q_{ip}Q_{jq}Q_{kr}Q_{l s}\eta_{pqrs}
\end{equation}
for all $Q_{ij}$ in the point group associated with a wallpaper group. 
Given our focus on mechanics, it is convenient to express the relations between symmetric stress and strain-rates in \citet{voigt:1910} form~\cite[see also][\S 7.1.1]{anandgovindjee}: 
\begin{equation}\label{eq:voigt}
\begin{bmatrix}
\sigma_{11} \\
\sigma_{22} \\
\sigma_{12}
\end{bmatrix}
=
\begin{bmatrix}
\mathrm{\eta}_{1111} & \mathrm{\eta}_{1122} & \mathrm{\eta}_{1112} \\
\mathrm{\eta}_{2211} & \mathrm{\eta}_{2222} & \mathrm{\eta}_{2212} \\
\mathrm{\eta}_{1211} & \mathrm{\eta}_{1222} & \mathrm{\eta}_{1212}
\end{bmatrix}
\begin{bmatrix}
\phantom{2}d_{11} \\
\phantom{2}d_{22} \\
2d_{12}
\end{bmatrix} \, .
\end{equation}
Anomalous behaviors manifest as asymmetries of the off-diagonal components of the ${3 \times 3}$ matrix in Eq.~\eqref{eq:voigt}, i.e., when $\eta_{ijk l}\neq \eta_{kl ij}$.

Analogous to the second-order tensors, we may analyze the representations of fourth-order tensors by first considering reflection and rotation symmetries independently.
To begin, 2D materials lacking both reflection and rotation symmetries, viz.\ p1~(o), and  materials with only two-fold rotational symmetry, viz.\ p2~(2222), have all 9 coefficients unconstrained and permit anisotropic anomalies involving all off-diagonal elements.

Now consider a reflection symmetry ${Q} = B_0$. Applying Eq.~\eqref{eq:4thorder-transform}, any component of $\eta_{ijkl}$ in which a 2 index appears an odd number of times must vanish. That is,
$\eta_{1112} = \eta_{1211} = \eta_{1222} = \eta_{2212} = 0$,
eliminating several scenarios involving major symmetry breaking.  The case of ${Q}=B_\pi$ provides the same net result, though with the reasoning using the appearance of the 1 index an odd number of times.
However, components with a 2 (respectively a 1) index appearing an even number of times, such as $\eta_{1122}$ and $\eta_{2211}$, remain unconstrained and can still exhibit major symmetry breaking. Thus, anomalous responses may arise in wallpaper groups:
pm~(**), pg~(XX), cm~(*X), p2mm~(*2222), p2mg~(22*), p2gg~(22X), and c2mm~(2*22).

Consider now the wallpaper group p4 (442) with rotational symmetries that are integer multiples of $\theta = \pi/2$, but with no reflection symmetry.
Applying the invariance requirement in Eq.~\eqref{eq:4thorder-transform} 
provides the relations that $\eta_{1111}=\eta_{2222}$, $\eta_{1122}=\eta_{2211}$, $\eta_{1112}=-\eta_{2212}$, and $\eta_{1211}=-\eta_{1222}$, while retaining the possibility of anomalous behavior due to the normal to shear couplings.  In total there are 5 independent elements to the tensor. 
However, introducing a reflection symmetry $B_0$, as in the wallpaper groups p4mm~(*442) and p4gm~(4*2), enforces $\eta_{1112} = \eta_{2212} = 0$, thereby reducing the number of independent coefficients to three and allowing no anomalous behavior.

Consider now the wallpaper groups that possess rotational symmetries by integer multiples of $\pi/3$, as in p3~(333) and p6~(632).
Analyzing these transformations via Eq.~\eqref{eq:4thorder-transform} can become cumbersome due to the non-zero off-diagonal entries in the rotation matrices. To address this, we adopt a complex variable transformation technique introduced by Khatkevich \cite{khatkevich:62} (see also~\cite[\S 10]{landau.lifshitz-elas:86} and Appendix Sections \ref{sec:c3} and \ref{sec:c34th} for derivational details).
This analysis restores all major symmetries, except those involving $\eta_{1112}$ and $\eta_{2212}$.
In fact, we find that
$\eta_{1112} = -\eta_{1211} = -\eta_{2212} = \eta_{1222}$,
indicating major symmetry breaking.
Furthermore, the rotational symmetries impose
$\eta_{2222} = \eta_{1111}$,
corresponding to an isotropic response of the normal stresses to the corresponding normal strains, and an additional relation 
$\eta_{1212} = \tfrac{1}{2}(\eta_{1111} - \eta_{1122})$.
Introducing the notations
$\eta_{1212} = \mu$,
$\eta_{1111} = \lambda + 2 \mu $, and
$\eta_{1112} = \mu^{\mathrm{o}}$,
one recovers the well-known two-dimensional Newtonian fluid model with $\lambda$ and $\mu$ related to the bulk and shear viscosities, respectively, augmented by an additional odd viscosity $\mu^\mathrm{o}$—just as in \emph{isotropic} 2D systems breaking both time-reversal symmetry and parity~\cite{Avr98,Banerjee2017,kranthi-odd-viscosity, Soni2019}. 

Lastly, the presence of any additional reflection symmetry involving either $B_0$ or $B_\pi$, as in p3m1~(*333), p31m~(3*3) and p6mm~(*632),  eliminates the odd viscosity, and restores a response identical to that of an isotropic system with preserved time-reversal symmetry and no anomalies.
A full list of the representation theorems for $4^\mathrm{th}$ order tensors for all wallpaper groups is enumerated in Column IV of Table~\ref{t:wallpaper_tensor}.

\vspace{0.1in}
\noindent\textbf{\textit{Discussion: Twisted bilayer graphene and knitted fabrics.}} 
Without invoking the microscopic origins of emergent behavior, we conclude with a discussion of the implications of the representation theorems for second- and fourth-order tensors in two distinct classes of two-dimensional condensed matter systems: twisted bilayer graphene and knitted fabrics. 

Twisted bilayer graphene systems are a class of moir\'{e} materials formed by stacking two graphene sheets with a relative twist. When aligned at a so-called “magic angle,” they have been shown to exhibit unconventional superconductivity~\cite{cao2018unconventional}. At certain commensurate twist angles, these systems form superlattices that can be classified using the wallpaper groups listed in Table~\ref{t:wallpaper_tensor}. 
These commensurate angles can be identified from a rational fraction corresponding to a pair of coprime integers $(m,n)$ (see Ref.~\cite{mele2010}), and are given by 
\begin{equation}
\cos(\theta(m,n)) = \frac{1}{2}\frac{m^2+n^2+4mn} {m^2+n^2+mn} \, .
\end{equation}
In the superlattice structure, when atoms from different monolayers coincide, these atom pairs are always from the same sublattice. 
The resulting superlattices then fall into two families, distinguished by a sublattice exchange (SE) symmetry~\cite{mele2010}: 
SE-even, when these atom pairs come from both sublattices and SE-odd, when they originate from only one sublattice.
This distinction corresponds to the condition: $(m - n) \bmod 3 = 0$ for SE-even structures, and otherwise SE-odd~\cite{mele2010}.

SE-even structures contain centers of six-fold rotation but lack reflection symmetry, corresponding to the wallpaper
group p6 (632). SE-odd structures also lack reflection symmetry but possess centers of three-fold rotation, corresponding to the wallpaper group p3 (333). These symmetries allow for the observation of anomalous material property tensors (see Table~\ref{t:wallpaper_tensor}), for example, fourth-order odd viscosity tensors---even in the absence of a magnetic field. The exceptions are the twist angles $\theta = 0^\circ$, $60^\circ$, and $120^\circ$, at which reflection symmetries are restored. At these angles, the resulting superlattice structures correspond to the wallpaper groups p6mm (*632), p3m1 (*333), and p6mm (*632), respectively, and exhibit no anomalous behavior as seen from Table~\ref{t:wallpaper_tensor}. 

For any irrational or non-commensurate twist angle $\theta$, twisted bilayer graphene falls into the wallpaper group p1 (o). In this case, all nine coefficients of the viscosity tensor are unconstrained, allowing for anisotropic anomalies involving off-diagonal elements.
At the special angle $\theta = 30^\circ$, the bilayers form a dodecagonal quasicrystalline structure~\cite{yao2018}, whose representations and material behaviors lie outside the scope of this work.
All of the aforementioned conclusions also apply to second-order tensors governing diffusive and resistive behavior, such as electric currents under applied electric fields.

While twisted bilayer graphene superlattices predominantly exhibit wallpaper groups p3 (333), p6 (632) and their mirror-symmetric counterparts, knitted fabrics present an alternate system, one that can be designed to realize nearly all of the wallpaper groups listed in Table~\ref{t:wallpaper_tensor}. Furthermore, the mechanical behavior of these materials can be tuned by varying the topology of the underlying stitch \cite{Singal2023}. The 2D projections of both a fabric with uniform knit stitches and a fabric with alternating rows of knits and purls both belong to wallpaper group p2mg (22*)~\cite{OKeeffe2022IsogonalEmbeddings, ITC_VolE}. Moreover, it is possible to design knitted fabrics conforming to the wallpaper group p4 (442) using a technique known in the knitting community as entrelac knitting. While less is known about structures conforming to p1 (o), p2 (2222), p3 (333) and p6 (632) wallpaper group symmetries, such patterns could be achieved by altering the basic stitches in combination with sewing techniques. This versatility offers a promising platform for exploring anomalous behaviors. 

Fabrics are solids, and at first glance, one might expect their elastic behavior described by the fourth-order elasticity tensor $C_{ijkl}$ to exhibit anomalies in any structure corresponding to mirror-symmetry-breaking wallpaper groups.
However, the existence of such anomalous behavior must be consistent with the second law of thermodynamics. For mechanical systems without external energy input, this law requires $\oint \sigma_{ij}\dot\epsilon_{ij} \,dt\ge 0$ for any closed deformation cycle, where $\epsilon_{ij}$ is the strain. 
In the case of elasticity, this condition is equivalent to a constraint on the anti-symmetric part of the elastic stiffness, viz., $\frac{1}{2} \oint \epsilon_{kl} (C_{ijkl} - C_{klij}) \dot\epsilon_{ij} \, dt \ge 0$.
Consider now a fabric with wallpaper group p6 (632) subjected to the strain cycle $\epsilon_{11}(t) = A\sin\omega t$, $\epsilon_{22}(t)= A\cos\omega t$, $\epsilon_{12}(t) = A\cos\omega t$, in the time interval $[0, 2\pi/\omega]$ for any infinitesimally small amplitude $A$. 
Given the representation theorem for p6 (632) listed in Table~\ref{t:wallpaper_tensor}, one observes that $\frac{1}{2}\oint   \epsilon_{kl} [C_{ijkl}-C_{klij}] \dot\epsilon_{ij}\, dt = - 4\pi A^2 \mu^\mathrm{o} \ge 0 $, where $\mu^\mathrm{o}$ is the odd elasticity coefficient \cite{scheibner2020odd}.  A flip of the algebraic sign of the shear strain yields the opposite result; thus
the inequality can only be satisfied if $\mu^{\mathrm{o}} = 0$, implying that odd elasticity must vanish. 
Thus, no odd elastic behavior is allowed without energy exchange during the deformation cycle---this result also rules out anomalous elasticity in all fabrics.

Nevertheless, it is known that fabrics exhibit pronounced stress-strain hysteresis (dissipation) \cite{Matsuo2009HysteresisKnitted,poincloux2018,dresselhaus.ea:25}. While the second law places restrictions on anomalous elastic responses, it imposes no restrictions on dissipative or path-dependent behaviors. As a result, anomalous behaviors such as odd inelasticity, may arise 
in the hysteretic regime. 
It is here that fabrics may offer an opportunity to explore anomalous responses beyond conventional elasticity, e.g., in cyclic excitation and in coupled systems such as fabrics endowed with yarn-level conductivity; additionally, one could explore odd elastic behaviors in twisted bilayer graphene superlattices in an electromagnetic field. While the ideas presented in this section remain speculative, it is our hope that they inspire new experiments in search of novel major symmetry breaking anomalous behaviors. \\
\\

\begin{acknowledgments}
EJD thanks Anton Souslov, Elisabetta Matsumoto, Michael Dimitriyev and Sonia Mahmoudi for helpful discussions. EJD, SG, and KKM acknowledge support from the National Science Foundation (Grant No.\ CMMI-2344589). Any opinions, findings, and conclusions expressed in this publication are those of the authors and do not necessarily reflect the views of the NSF. KKM also acknowledges partial support from the Director, Office of Science, Office of Basic Energy Sciences, of the U.S. Department of Energy under contract No. DEAC02-05CH11231. 
\end{acknowledgments}

\bibliography{wallp}

\renewcommand{\theequation}{\thesection.\arabic{equation}}
\renewcommand{\thefigure}{\thesection.\arabic{figure}}
\renewcommand{\thesection}{\Alph{section}}
\renewcommand{\thesubsection}{\thesection.\arabic{subsection}}

\makeatletter
\renewcommand{\p@subsection}{}  
\renewcommand{\p@subsubsection}{}
\makeatother

\clearpage
\begin{center}
\textbf{Appendices} \\
\vspace{0.05in}
\end{center}

\setcounter{equation}{0}
\setcounter{footnote}{0}
\setcounter{section}{0}
\setcounter{figure}{0}

\section{Symmetries: Translations, Rotations and Reflections}

In two dimensions, the Euclidean plane is invariant under the Euclidean group $E_2$.  
Following \citet{armstrong:1988} (see also \citet{sasse:2020}),
this is the group of all isometries of the plane $\IR^2$:
$$ E_2 = \{ g(x) ~|~ \Vert g(x) - g(y) \Vert = \Vert x - y \Vert \quad \forall_{x,y \in \IR} \} \,.$$
The functions $g(\cdot)$ can all be represented by a vector in the plane and an element of $O(2)$, the orthogonal group over $\IR^2$, as
\begin{equation}\label{eq:iso} g(x) = v + Mx\,,\end{equation}
where $v\in\IR^2$ and $M\in O(2)$.  Thus the elements of $E_2$ can be considered as ordered pairs of the form $(v,M)$ with the group operation defined via
$$(v,M)(w,N) = (v+Mw,MN)\,.$$
The elements of the Euclidean group can be classified as translations, rotations, and reflections (pure and glide).  
\begin{itemize}
\item\textbf{Translations} occur for group elements of the form (v,$\mathbb{I}$), with $\mathbb{I}$ being the identity tensor.

\item\textbf{Rotations} about a point $c$ occur for elements of the form $(v,M) = (c,\mathbb{I})(0,M)(-c,\mathbb{I})$ with $\det{M} = +1$; note $v =(\mathbb{I}-M)c$.  

\item\textbf{Pure reflections} occur when $Mv=-v$ and $\det{M}=-1$. In this case the group elements are of the form $(v,M)=(v/2,\mathbb{I})(0,M)(-v/2,\mathbb{I})$.  The plane is reflected about a line orthogonal to $v$ and shifted from the origin by $v/2$.

\item\textbf{Glide reflections} occur when $Mv\not= -v$ and $\det{M}=-1$. In this case the group elements are of the form $(v,M)=(b,I)(a,M)$ where $b = v - a$, $a = (w\otimes w) v$ and $w = (v-Mv)/\Vert v-Mv\Vert$.  This implies a reflection of the plane about a line parallel to $b$ (or orthogonal to $a$) and is displaced from the origin by $a/2$, followed by a translation $(b,I)$ parallel to $b$. (Note $a\cdot b = 0$).
\end{itemize}

The wallpaper groups $G$ are subgroups of $E_2$, where the translation elements are defined by a two dimensional basis, $v = n\, a + m\, b$ for all ${n,m\in\IZ}$ and some $a,b \in \IR^2$.  The orthogonal elements of $G$ are restricted to be in a finite subgroup of $O(2)$ which is termed a point group.

As enumerated by \cite{schwarzenberger:1974, schattschneider:1978,armstrong:1988} among many others, there are 17 unique wallpaper groups.  Our interest lies in the implications of the groups on the linear coarse-grained properties of two-dimensional materials, i.e.\ continuum properties as represented by tensors of varying order.  Thus our interest lies in the point groups associated to each wallpaper group.  Without detailing the nomenclature, distinct point groups are as listed in Column III of Table~\ref{t:wallpaper_tensor}~\citep[Chapter 26]{armstrong:1988}.  Wallpaper groups that share the same point group are aggregated on the same line.\footnote{ Note that for p3m1~(*333) and p31m~(3*3), both of which share the point group $D_3$, we provide two equivalent sets of reflection elements as is commonly found in the literature.}
The differentiation between groups on the same line in the table is related to the number of permutations of pure and glide reflections that can occur. Also, note that $-\mathbb{I}\equiv A_\pi$ in two dimensions (unlike in three dimensions).

\section{Representations for second-order tensors}\label{app:B}
For materials that posses symmetries in the various wallpaper groups, there is a reduction in the number of independent material constants.  In this section, we will analyze the implications of the wallpaper groups on the second-order tensors related to transport properties such as resistivity or diffusivity that form the representations summarized in Column V of Table~\ref{t:wallpaper_tensor}. As in the main text, we will consider the constitutive relation relating the electric field ${E}$ and the associated current ${J}$ of the form ${E} = {\rho} {J}$, with the tensor components:
$$
[\rho] = \begin{bmatrix} \rho_{11} & \rho_{12} \\ \rho_{21} & \rho_{22} \end{bmatrix} 
\,.
$$

The analysis for second-order tensors can be performed by applying Eq.~\eqref{eq:2nd-transformation} for each point group.   It is convenient to recall that a second-order tensor is a multi-linear operator over 2 copies of a vector space \cite{schutz:80} and that the components are given by
$$
\rho_{ij} = \rho(\be_i,\be_j)\,.
$$
If $Q$ is a symmetry transformation of a material, then \eqref{eq:2nd-transformation} can be equivalently expressed as
\begin{equation}
\rho(\be_i,\be_j) = \rho(Q\be_i,Q\be_j)\,.
    \label{e:symtran}
\end{equation}

\subsection{p1 (o): $C_1$}

For materials with the point group $C_1$, there are 4 possible resistivities.  No reductions are possible and therefore major symmetry is broken.

\subsection{p2 (2222): $C_2$}

For materials with the point group of $C_2$, we have the inter-relations that $\be_1\rightarrow -\be_1$ and $\be_2 \rightarrow -\be_2$.  In this setting no reductions are possible as in the p1 (o) case and there are possibly 4 unique constants again with no major symmetry.

\subsection{pm (**), pg (XX), cm (*X): $D_1$}

For materials with the point group $D_1$, we have the inter-relations 
\begin{align*}
B_0:&&&\be_1 \rightarrow \be_1  &&\be_2\rightarrow -\be_2\,.
\end{align*}
Thus,
$$
[\rho] = \begin{bmatrix}
\rho_{11} & 0 \\
0 & \rho_{22} 
\end{bmatrix}\,.
$$
This can also be readily seen by applying \eqref{e:symtran} with $Q = B_0$ corresponding to reflection about the $x$-axis. Overall there are possibly 2 unique material constants and major symmetry is preserved.

\subsection{p2mm (*2222), p2mg (22*), p2gg (22X), c2mm (2*22): $D_2$}
For materials with the point group $D_2$, we have the inter-relations
\begin{align*}
B_0:&&&\be_1 \rightarrow \be_1  &&\be_2\rightarrow -\be_2\\
B_\pi:&&&\be_1 \rightarrow -\be_1  &&\be_2\rightarrow \be_2\,.
\end{align*}
These four space groups yield identical results to pm (**), pg (XX), and cm (*X), again giving possibly 2 unique material constants:
$$
[\rho] = \begin{bmatrix}
\rho_{11} & 0 \\
0 & \rho_{22} 
\end{bmatrix}
$$
and the preservation of major symmetry.

\subsection{p4 (442): $C_4$}
For materials with point group $C_4$, we have the inter-relations
\begin{align*}
A_{\pi/2}:&&&\be_1 \rightarrow \be_2  &&\be_2\rightarrow -\be_1\\
A_{3\pi/2}:&&&\be_1 \rightarrow -\be_2  &&\be_2\rightarrow \be_1\,.
\end{align*}
These imply the following structure to the resistivity tensor:
$$
[\rho] =    \begin{bmatrix}
\rho_{11} & \rho_{12} \\
-\rho_{12} & \rho_{11} 
\end{bmatrix}   \,.
$$
We find 2 unique material constants and the anomaly of oddity.

\subsection{p4mm (*442), p4gm (4*2): $D_4$}

For materials with point group $D_4$, we have the inter-relations
\begin{align*}
A_{\pi/2}:&&&\be_1 \rightarrow \be_2  &&\be_2\rightarrow -\be_1\\
A_{3\pi/2}:&&&\be_1 \rightarrow -\be_2  &&\be_2\rightarrow \be_1\\
B_0:&&&\be_1 \rightarrow \be_1  &&\be_2\rightarrow -\be_2\\
B_{\pi/2}:&&&\be_1 \rightarrow\be_2 && \be_2\rightarrow \be_1\\
B_\pi:&&&\be_1 \rightarrow -\be_1  &&\be_2\rightarrow \be_2\\
B_{3\pi/2}&&&\be_1 \rightarrow -\be_2 && \be_2\rightarrow -\be_1
\,.
\end{align*}
Some of these transformations correspond to the previous cases. The two additional symmetries associated with $B_{\pi/2}$ and $B_{3\pi/2}$ simply yield $\rho_{11}= \rho_{22}$ and the major symmetry $\rho_{12} = \rho_{21}$, which when combined with the other symmetries lead to vanishing off-diagonal components. Overall, these wallpaper groups follow an isotropic structure for the resistivity tensor:
$$
[\rho] = \begin{bmatrix}
\rho_{11} & 0 \\
0 & \rho_{11} 
\end{bmatrix}\,,
$$
where we find 1 unique material constant and no anomalies.

\subsection{p3 (333): $C_3$}\label{sec:c3}
For materials with point group $C_3$, we have the inter-relations
\begin{align*}
A_{2\pi/3}:&&&\be_1 \rightarrow -\frac{1}{2}\be_1+\frac{\sqrt{3}}{2}\be_2  
                   &&\be_2 \rightarrow -\frac{\sqrt{3}}{2}\be_1 -\frac{1}{2}\be_2\\
A_{4\pi/3}:&&&\be_1 \rightarrow -\frac{1}{2}\be_1-\frac{\sqrt{3}}{2}\be_2  
                   &&\be_2 \rightarrow \frac{\sqrt{3}}{2}\be_1-\frac{1}{2}\be_2
\,.
\end{align*}

The analysis for this point group and other point groups with rotations/reflections that are integer multiples of $\pi/3$ is somewhat cumbersome.   However, the technique found in \cite{khatkevich:62} \cite[see also,][\S 10]{landau.lifshitz-elas:86} using a complex variable transformation can assist in reducing some of the tedium associated with the computation.\footnote{This technique is strongly reminiscent of the analytic tools used to study commensurate angles in twisted bilayer graphene \cite{mele2010,lopes.ea:07}.} The essence of the method is to perform a coordinate transformation $x_i \rightarrow \xi_i$ where $\xi_i$ are complex-valued scalars.  This generates a new complex-valued basis in which the components of the resistivity tensor are quite simple in form for these point groups.   This result is then transformed back to the real-valued basis to yield the final result.  Since descriptions of this technique are not easily found in contemporary literature, we describe the analysis steps in some detail. While the benefit for second-order tensors may be modest, this complex transformation technique proves invaluable for analyzing higher-order tensors that contain a larger number of independent components, for instance fourth-order tensors as we shall see later.
\smallskip
\paragraph{Complex transformation:} 
We introduce the change of variables
$$
\begin{pmatrix} \xi_1 \\ \xi_2 \end{pmatrix} =
\underbrace{\begin{pmatrix} 1 & \mathfrak{i} \\ 1 & -\mathfrak{i} \end{pmatrix}}_{C}
\begin{pmatrix} x_1 \\ x_2 \end{pmatrix}
$$
with inverse
$$
\begin{pmatrix} x_1 \\ x_2 \end{pmatrix}
 =
\frac{1}{2}\begin{pmatrix} 1 &1 \\ -\mathfrak{i} & \mathfrak{i} \end{pmatrix}
\begin{pmatrix} \xi_1 \\ \xi_2 \end{pmatrix}\,,
$$
where $\mathfrak{i}=\sqrt{-1}$.  The natural basis vectors in the $\xi_i$ coordinates are given
by $\frac{\partial}{\partial\xi_i} = \frac{\partial x_j}{\partial \xi_i} \frac{\partial}{\partial x_j}$.  If we denote the normalized basis vectors in the complex space as $\ba_i$, then
$$
\begin{aligned}
\ba_1 &= \frac{1}{\sqrt{2}}(\be_1 - \mathfrak{i}\be_2) \\
\ba_2 &= \frac{1}{\sqrt{2}}(\be_1 + \mathfrak{i}\be_2)
\end{aligned}
$$
with inverse
$$
\begin{aligned}
\be_1 &= \frac{1}{\sqrt{2}}(\ba_1 + \ba_2) \\
\be_2 &= \frac{\mathfrak{i}}{\sqrt{2}}(\ba_1 - \ba_2)\,.
\end{aligned}
$$
The convenience afforded by this transformation becomes apparent when one notes that
a rotation $A_\theta$ of the physical plane induces the following transformation in the complex space: 
\begin{equation}
\begin{pmatrix} \xi_1 \\ \xi_2 \end{pmatrix} \rightarrow C A_\theta C^{-1}
 \begin{pmatrix} \xi_1 \\ \xi_2 \end{pmatrix} = \begin{pmatrix} e^{\mathfrak{i}\theta} & 0 \\
 0 & e^{-\mathfrak{i}\theta} \end{pmatrix} \begin{pmatrix} \xi_1 \\ \xi_2 \end{pmatrix}\,. \label{e:rotc} 
\end{equation}
A reflection $B_\alpha$ of the plane induces the transformation in the complex space:
\begin{equation}
\begin{pmatrix} \xi_1 \\ \xi_2 \end{pmatrix} \rightarrow C B_\alpha C^{-1}
 \begin{pmatrix} \xi_1 \\ \xi_2 \end{pmatrix} = \begin{pmatrix} 0&e^{\mathfrak{i}\alpha} \\
  e^{-\mathfrak{i}\alpha}&0 \end{pmatrix} \begin{pmatrix} \xi_1 \\ \xi_2 \end{pmatrix}\,.
\label{e:refc}
\end{equation}
 
The tangent map of these transformations will map the basis vectors $\ba_i$ according to
\cite{schutz:80,marsden.hughes:93}:
$$
\text{Rotation:} \quad
\ba_1 \rightarrow e^{\mathfrak{i}\theta} \ba_1
\qquad
\ba_2 \rightarrow e^{-\mathfrak{i}\theta} \ba_2
$$
and
$$\text{{Reflection:}} \quad
\ba_1 \rightarrow e^{-\mathfrak{i}\alpha} \ba_2
\qquad
\ba_2 \rightarrow e^{\mathfrak{i}\alpha} \ba_1\,.
$$
\smallskip
\paragraph{Complex-valued components:}
Let us denote the complex-valued resistivity tensor components as 
$$
[\rho'] = \begin{bmatrix} \rho_{1'1'} & \rho_{1'2'} \\ \rho_{2'1'} & \rho_{2'2'} \end{bmatrix} \,.
$$
where the $(\cdot)'$ notation is used to designate components in the complex-valued basis. 
The individual, possibly complex-valued, components are computed as:
$$
\rho_{i'j'} = \rho(\ba_i,\ba_j)\,.
$$
As before the tensor $\rho$ should be taken to be a multi-linear operator.

\smallskip
\paragraph{Applications to material symmetry:}
Consider now the use of symmetry requirement \eqref{e:symtran} for a rotation. In the complex basis,
\begin{align*}
    \rho(\ba_1,\ba_1) = \rho(e^{\mathfrak{i}\theta}\ba_1,e^{\mathfrak{i}\theta}\ba_1)
    = e^{i2\theta}\rho(\ba_1,\ba_1)\,.
\end{align*}
For symmetry transformations $A_{2\pi/3}$ and $A_{4\pi/3}$, it is clear then that $\rho_{1'1'}$ must be zero for the relation to hold. Similar reasoning hold for $\rho_{2'2'}$.  For the off-diagonal terms, one has for example
\begin{align*}
    \rho(\ba_1,\ba_2) = \rho(e^{\mathfrak{i}\theta}\ba_1,e^{-\mathfrak{i}\theta}\ba_2)
    = e^{i(\theta-\theta)}\rho(\ba_1,\ba_1)=\rho(\ba_1,\ba_2)\,.
\end{align*}
Thus,
one finds that the complex-valued components in the $\ba_i$ basis reduce,
other than for rotational symmetries of integer multiples of $\pi$,
to the form
$$
[\rho'] = \begin{bmatrix} 0 & \rho_{1'2'} \\ \rho_{2'1'} & 0 \end{bmatrix} \,.
$$

\paragraph{Real-valued components:} From the non-zero complex-valued components of the form above, the real-valued components can be obtained using the linear map properties as \begin{widetext}
$$
\begin{alignedat}{4}
\rho_{11} &= \rho(\be_1,\be_1) &&= \frac{1}{2}\rho(\ba_1+\ba_2,\ba_1+\ba_2) &&= \frac{1}{2}( \rho_{1'2'}+\rho_{2'1'}) \\
\rho_{22} &= \rho(\be_2,\be_2) &&= \frac{\mathfrak{i}^2}{2}\rho(\ba_1-\ba_2,\ba_1-\ba_2) &&= \frac{1}{2}( \rho_{1'2'}+\rho_{2'1'}) &&=\phantom{-}\rho_{11} \\
\rho_{12} &=\rho(\be_1,\be_2) &&= \frac{\mathfrak{i}}{2}\rho(\ba_1+\ba_2,\ba_1-\ba_2) &&=\frac{\mathfrak{i}}{2}(\rho_{2'1'}-\rho_{1'2'})\\
\rho_{21} &=\rho(\be_2,\be_1) &&= \frac{\mathfrak{i}}{2}\rho(\ba_1-\ba_2,\ba_1+\ba_2) &&=\frac{\mathfrak{i}}{2}(\rho_{1'2'}-\rho_{2'1'})&&=-\rho_{12}\,,
\end{alignedat}
$$
\end{widetext}
yielding the following structure for the components (in the physical basis)
$$
[\rho] = \begin{bmatrix}
\rho_{11} & \rho_{12} \\
-\rho_{12} & \rho_{11} 
\end{bmatrix}\,.
$$
with 2 unique material constants and the anomaly of oddity.

Observe that one can equivalently determine the same relations by expanding the right-hand side 
and matching terms in the identity
$$
T_{ij}\be^i\otimes\be^j = T_{i'j'}\ba^i\otimes\ba^j\,,
$$
where it is noted that the dual vectors as we have set them  up have the relations $\be^i = \be_i$ and $\ba^i = \overline{\ba_i}$.

\subsection{p3m1 (*333), p31m (3*3): $D_3$}
For point group $D_3$ the important additional symmetry group elements are
\begin{align*}
B_\pi:&&&\ba_1 \rightarrow -\ba_2  &&\ba_2\rightarrow -\ba_1
\end{align*}
or equivalently
\begin{align*}
B_0:&&&\ba_1 \rightarrow \ba_2  &&\ba_2\rightarrow \ba_1
\,.
\end{align*}
Either one implies $\rho_{1'2'}=\rho_{2'1'}$, yielding
$$
[\rho] = \begin{bmatrix}
\rho_{11} & 0 \\
0 & \rho_{11} 
\end{bmatrix}\,.
$$
and 1 unique constant and no anomalies.

\subsection{p6 (632): $C_6$}
The symmetry elements of $C_6$ result in the same conclusion as with point group $C_3$:
$$
[\rho] = \begin{bmatrix}
\rho_{11} & \rho_{12} \\
-\rho_{12} & \rho_{11} 
\end{bmatrix}\,.
$$
with 2 unique constants and the anomaly of oddity.   

\subsection{p6mm (*632): $D_6$ }
Lastly, the symmetry elements of $D_6$ yield similar results to the analysis of $D_3$:
$$
[\rho] = \begin{bmatrix}
\rho_{11} & 0 \\
0 & \rho_{11} 
\end{bmatrix}\,.
$$
with 1 unique constant and no anomalies.

\section{Representations for fourth-order tensors}
Here we examine the implications of the wallpaper symmetry groups on the fourth-order tensors for material properties concerning viscosity or elasticity moduli (in two dimensions) without the assumption of major symmetry.  Focusing on the viscosity tensor, throughout, we will assume minor symmetry of $\eta_{ijkl}$, i.e., $\eta_{ijkl}=\eta_{jikl}=\eta_{ijlk}$, as one would have for non-polar media.  However, in general, $\eta_{ijkl}\not=\eta_{klij}$, as it is our objective to identify which wallpaper groups permit this major symmetry to be broken.  In this context the stress $\sigma_{ij}=\eta_{ijkl}d_{kl}$, where $d_{kl}$ is the symmetric gradient of the velocity field.  As such we work with the components of $\eta_{ijkl}$ in the form 
$$
[\eta] = \begin{bmatrix}
\eta_{1111} & \eta_{1122} & \eta_{1112} \\
\eta_{2211} & \eta_{2222} & \eta_{2212} \\
\eta_{1211} & \eta_{1222} & \eta_{1212}
\end{bmatrix}\,;
$$
see also \eqref{eq:voigt}. It is convenient to recall that a fourth-order tensor is a multi-linear operator over 4 copies of a vector space \cite{schutz:80} and that the components are given by
\begin{equation}
\eta_{ijkl} = \eta(\be_i,\be_j,\be_k,\be_l)\,.\label{e:fourthcomp}
\end{equation}
Thus invariance requirement \eqref{eq:4thorder-transform} can alternately be written as
\begin{equation}
    \eta(\be_i,\be_j,\be_k,\be_l) = \eta(Q\be_i,Q\be_j,Q\be_k,Q\be_l)\,,\label{e:symtran4}
\end{equation}
where $Q$ is any symmetry element.

\subsection{p1 (o): $C_1$}

For materials in point group $C_1$, there are 9 possible viscous coefficients.  No reductions are possible, allowing major symmetry breaking among all off diagonal components.

\subsection{p2 (2222): $C_2$}

For materials with point group $C_2$, we have the inter-relations that $\be_1\rightarrow -\be_1$ and $\be_2 \rightarrow -\be_2$.  In this setting no reductions are possible and there are possibly 9 unique constants allowing major symmetry to be broken among all off diagonal components.
Also, note that since point group elements $\mathbb{I}$ and $-\mathbb{I}$ have no implications on the tensor representation, in what follows we do not explicitly discuss them in our analysis.

\subsection{pm (**), pg (XX), cm (*X): $D_1$}

For materials in point group $D_1$, we have the inter-relations 
\begin{align*}
B_0:&&&\be_1 \rightarrow \be_1  &&\be_2\rightarrow -\be_2\,.
\end{align*}
Thus, it can be readily seen from \eqref{eq:4thorder-transform} or \eqref{e:symtran4} that
$$
[\eta] = \begin{bmatrix}
\eta_{1111} & \eta_{1122} & 0 \\
\eta_{2211} & \eta_{2222} & 0 \\
0 & 0 & \eta_{1212}
\end{bmatrix}\,,
$$
and there are possibly 5 unique material constants allowing major symmetry breaking.

\subsection{p2mm (*2222), p2mg (22*), p2gg (22X), c2mm (2*22): $D_2$}
For materials having point group $D_2$, we have the inter-relations
\begin{align*}
B_0:&&&\be_1 \rightarrow \be_1  &&\be_2\rightarrow -\be_2\\
B_\pi:&&&\be_1 \rightarrow -\be_1  &&\be_2\rightarrow \be_2\,.
\end{align*}
From the perspective of fourth-order tensor representations with minor symmetry, the invariance requirements of $D_2$ match those of $D_1$.  There are 5 unique material constants and anomalous behavior is possible.

\subsection{p4 (442): $C_4$}
For materials with  point group $C_4$, we have the inter-relations
\begin{align*}
A_{\pi/2}:&&&\be_1 \rightarrow \be_2  &&\be_2\rightarrow -\be_1\\
A_{3\pi/2}:&&&\be_1 \rightarrow -\be_2  &&\be_2\rightarrow \be_1\,.
\end{align*}
These imply the following structure to the viscosity tensor
$$
[\eta
] = \begin{bmatrix}
\eta_{1111} & \eta_{1122} & \eta_{1112} \\
\eta_{1122} & \eta_{1111} & -\eta_{1112} \\
\eta_{1211} & -\eta_{1211} & \eta_{1212}
\end{bmatrix}\,,
$$
which can be readily seen from application of \eqref{eq:4thorder-transform} or \eqref{e:symtran4}. Thus, we find 5 unique constants in this case, and major symmetry is allowed to be broken through the shear-normal coupling.

\subsection{p4mm (*442), p4gm (4*2): $D_4$}

For materials with point group $D_4$, we have the inter-relations
\begin{align*}
A_{\pi/2}:&&&\be_1 \rightarrow \be_2  &&\be_2\rightarrow -\be_1\\
A_{3\pi/2}:&&&\be_1 \rightarrow -\be_2  &&\be_2\rightarrow \be_1\\
B_0:&&&\be_1 \rightarrow \be_1  &&\be_2\rightarrow -\be_2\\
B_{\pi/2}:&&&\be_1 \rightarrow\be_2 && \be_2\rightarrow \be_1\\
B_\pi:&&&\be_1 \rightarrow -\be_1  &&\be_2\rightarrow \be_2\\
B_{3\pi/2}&&&\be_1 \rightarrow -\be_2 && \be_2\rightarrow -\be_1
\,.
\end{align*}
Following earlier analysis for the cases $D_2$ and $C_4$, 
the above inter-relations imply the following structure 
$$
[\eta] = \begin{bmatrix}
\eta_{1111} & \eta_{1122} & 0 \\
\eta_{1122} & \eta_{1111} & 0 \\
0 & 0 & \eta_{1212}
\end{bmatrix}\,.
$$
In this case, we find 3 unique moduli and no anomalous behavior is allowed.

\subsection{p3 (333): $C_3$}\label{sec:c34th}

For materials with point group $C_3$, we encounter the same technical situation as seen with the case of second-order tensors in Sec.~\ref{sec:c3}.  As such we extend the analysis tools developed to the case of fourth-order tensors.  In particular we note that with the complex mapping of Sec.~\ref{sec:c3}, we can exploit the following relation to determine the symmetry-induced structure of fourth-order tensors:
$$
\text{Complex components:}\quad \eta_{i'j'k'l'} = \eta(\ba_i,\ba_j,\ba_k,\ba_l)\,.
$$
With this, a rotational symmetry where $\ba_1 \rightarrow e^{\mathfrak{i}\theta} \ba_1$
and $
\ba_2 \rightarrow e^{-\mathfrak{i}\theta} \ba_2$ would, for example, require
\begin{multline*}
\eta(\ba_1,\ba_1,\ba_1,\ba_1) = \eta(e^{\mathfrak{i}\theta}\ba_1,e^{\mathfrak{i}\theta}\ba_1,e^{\mathfrak{i}\theta}\ba_1,e^{\mathfrak{i}\theta}\ba_1) \\= e^{\mathfrak{i4}\theta}\eta(\ba_1,\ba_1,\ba_1,\ba_1)\,.
\end{multline*}
If, say, $\theta=2\pi/3$, this would require $\eta_{1'1'1'1'} = 0$.  Similar analysis can be made for reflections where $\ba_1 \rightarrow e^{-\mathfrak{i}\alpha} \ba_2$ and $
\ba_2 \rightarrow e^{\mathfrak{i}\alpha} \ba_1$.

\paragraph{Complex-valued components:}
Considering in more detail, the complex value components of the viscosity tensor are
$$
[\eta'] = \begin{bmatrix} \eta_{1'1'1'1'} & \eta_{1'1'2'2'} & \eta_{1'1'1'2'} \\ \eta_{2'2'1'1'} & \eta_{2'2'2'2'} & \eta_{2'2'1'2'} \\ \eta_{1'2'1'1'} & \eta_{1'2'2'2'} & \eta_{1'2'1'2'} \end{bmatrix}\,.
$$
For symmetry transformations $A_{2\pi/3}$ and $A_{4\pi/3}$ of $C_3$ one finds that the complex-valued components in the $\ba_i$ basis have the form
$$
[\eta'] = \begin{bmatrix} 0 & \eta_{1'1'2'2'} & 0 \\ \eta_{2'2'1'1'} & 0 & 0 \\ 0 & 0 & \eta_{1'2'1'2'} \end{bmatrix}\,.
$$

\paragraph{Real-valued components:} 
The real-valued components can be determined from \eqref{e:fourthcomp} and the relation of the physical basis to the complex basis:
\begin{widetext} 
$$
\begin{aligned}
\eta_{1111} &= \eta(\be_1,\be_1,\be_1,\be_1) = 
\frac{1}{4}\eta(\ba_1+\ba_2,\ba_1+\ba_2,\ba_1+\ba_2,\ba_1+\ba_2)=
\phantom{-}\frac{1}{4}( \eta_{1'1'2'2'}+\eta_{2'2'1'1'}) + \eta_{1'2'1'2'} \\
\eta_{2222} &= \eta(\be_2,\be_2,\be_2,\be_2) = 
\frac{\mathfrak{i}^4}{4}\eta(\ba_1-\ba_2,\ba_1-\ba_2,\ba_1-\ba_2,\ba_1-\ba_2)=
\phantom{-}\frac{1}{4}( \eta_{1'1'2'2'}+\eta_{2'2'1'1'}) + \eta_{1'2'1'2'} &&= \eta_{1111}\\
\eta_{1122} &= \eta(\be_1,\be_1,\be_2,\be_2) = 
\frac{\mathfrak{i}^2}{4}\eta(\ba_1+\ba_2,\ba_1+\ba_2,\ba_1-\ba_2,\ba_1-\ba_2)=
-\frac{1}{4}( \eta_{1'1'2'2'}+\eta_{2'2'1'1'}) + \eta_{1'2'1'2'} \\
\eta_{2211} &= \eta(\be_2,\be_2,\be_1,\be_1) = 
\frac{\mathfrak{i}^2}{4}\eta(\ba_1-\ba_2,\ba_1-\ba_2,\ba_1+\ba_2,\ba_1+\ba_2)=
-\frac{1}{4}( \eta_{1'1'2'2'}+\eta_{2'2'1'1'}) + \eta_{1'2'1'2'}&&= \eta_{1122}\\
\eta_{1212} &= \eta(\be_1,\be_2,\be_1,\be_2) = 
\frac{\mathfrak{i}^2}{4}\eta(\ba_1+\ba_2,\ba_1-\ba_2,\ba_1+\ba_2,\ba_1-\ba_2)=
\phantom{-}\frac{1}{4}( \eta_{1'1'2'2'}+\eta_{2'2'1'1'}) 
&&= \frac{1}{2}(\eta_{1111}-\eta_{1122})\\
\eta_{1112} &= \eta(\be_1,\be_1,\be_1,\be_2) =
\frac{\mathfrak{i}}{4}\eta(\ba_1+\ba_2,\ba_1+\ba_2,\ba_1+\ba_2,\ba_1-\ba_2)=
\phantom{-}\frac{\mathfrak{i}}{4}(\eta_{2'2'1'1'}-\eta_{1'1'2'2'}) \\
\eta_{2212} &= \eta(\be_2,\be_2,\be_1,\be_2) =
\frac{\mathfrak{i}^3}{4}\eta(\ba_1-\ba_2,\ba_1-\ba_2,\ba_1+\ba_2,\ba_1-\ba_2)=
-\frac{\mathfrak{i}}{4}(\eta_{2'2'1'1'}-\eta_{1'1'2'2'}) &&= -\eta_{1112}\\
\eta_{1211}&=
\eta(\be_1,\be_2,\be_1,\be_1) =
\frac{\mathfrak{i}}{4}\eta(\ba_1+\ba_2,\ba_1-\ba_2,\ba_1+\ba_2,\ba_1+\ba_2)=
-\frac{\mathfrak{i}}{4}(\eta_{2'2'1'1'}-\eta_{1'1'2'2'}) &&= -\eta_{1112}\\
\eta_{1222} &= \eta(\be_1,\be_2,\be_2,\be_2) =
\frac{\mathfrak{i}^3}{4}\eta(\ba_1+\ba_2,\ba_1-\ba_2,\ba_1-\ba_2,\ba_1-\ba_2)=
-\frac{\mathfrak{i}}{4}(\eta_{1'1'2'2'}-\eta_{2'2'1'1'}) &&= \phantom{-}\eta_{1112}\,.
\end{aligned}
$$
\end{widetext}
These relations yield the following structure, in the real basis, for the viscosity tensor: 
$$
[\eta] = \begin{bmatrix}
\eta_{1111} & \eta_{1122} & \eta_{1112} \\
\eta_{1122} & \eta_{1111} & -\eta_{1112} \\
-\eta_{1112} & \eta_{1112} & 
\frac{1}{2}(\eta_{1111}-\eta_{1122})
\end{bmatrix}
$$
with 3 unique constants.  Without loss of generality, we introduce $\eta_{1212} = \mu$, $\eta_{1111} = \lambda + 2\mu$,  and $\eta_{1112} =  \mu^\mathrm{o}$, then the viscosity tensor for point group $C_3$ is expressed as
$$
[\eta] = \begin{bmatrix}
\lambda + 2 \mu & \lambda  & \mu^\mathrm{o} \\
\lambda  & \lambda + 2 \mu & -\mu^\mathrm{o} \\
-\mu^\mathrm{o} & \mu^\mathrm{o} & 
\mu
\end{bmatrix}\,.
$$
Here, $\lambda$ and $\mu$ are related to the bulk and shear viscosities, respectively, while $\mu^o$ represents the odd viscosity resulting in the normal-shear coupling terms as in odd isotropic systems.

\subsection{p3m1 (*333), p31m (3*3): $D_3$}
For point group $D_3$ the additional symmetry group elements are either 
\begin{align*}
B_\pi:&&&\ba_1 \rightarrow -\ba_2  &&\ba_2\rightarrow -\ba_1
\,.
\end{align*}
or
\begin{align*}
B_0:&&&\ba_1 \rightarrow \ba_2  &&\ba_2\rightarrow \ba_1
\,.
\end{align*}
Either case
implies $\eta_{1'1'2'2'}=\eta_{2'2'1'1'}$, removing the oddities that arise with the $C_3$ group. Materials with $D_3$ point symmetry follow the structure
\begin{multline*}
[\eta] = \begin{bmatrix}
\eta_{1111} & \eta_{1122} & 0 \\
\eta_{1122} & \eta_{1111} & 0 \\
0 & 0 & \frac{1}{2}(\eta_{1111}-\eta_{1122}) 
\end{bmatrix} \\= \begin{bmatrix}
\lambda + 2 \mu & \lambda  & 0 \\
\lambda  & \lambda + 2 \mu & 0 \\
0 & 0 & 
\mu
\end{bmatrix}\, .
\end{multline*}
No anomalous behavior is allowed and the system is seen to respond exactly like an isotropic material, having 2 unique constants.

\subsection{p6 (632): $C_6$}
The symmetry elements of p6 result in the same conclusion as with wallpaper group p3:
\begin{multline*}
[\eta] = \begin{bmatrix}
\eta_{1111} & \eta_{1122} & \eta_{1112} \\
\eta_{1122} & \eta_{1111} & -\eta_{1112} \\
-\eta_{1112} & \eta_{1112} & 
\frac{1}{2}(\eta_{1111}-\eta_{1122})
\end{bmatrix}\\
= \begin{bmatrix}
\lambda + 2 \mu & \lambda  & \mu^\mathrm{o} \\
\lambda  & \lambda + 2 \mu & -\mu^\mathrm{o} \\
-\mu^\mathrm{o} & \mu^\mathrm{o} & 
\mu
\end{bmatrix}
\end{multline*}
with 3 unique constants.  Note the so-called odd-viscous structure in the normal-shear coupling terms.

\subsection{p6mm (*632): $D_6$}
Point group $D_6$, yields the same results as point group $D_3$:
\begin{multline*}
[\eta] = \begin{bmatrix}
\eta_{1111} & \eta_{1122} & 0 \\
\eta_{1122} & \eta_{1111} & 0 \\
0 & 0 & \frac{1}{2}(\eta_{1111}-\eta_{1122}) 
\end{bmatrix} \\= \begin{bmatrix}
\lambda + 2 \mu & \lambda  & 0 \\
\lambda  & \lambda + 2 \mu & 0 \\
0 & 0 & 
\mu
\end{bmatrix}\, .
\end{multline*}
No anomalous behavior is allowed and the system is seen to respond exactly like an isotropic material, having two unique constants.

\end{document}